# AI-based Personalization and Trust in Digital Finance

Vijaya Kanaparthi
Microsoft
Northlake, Texas, USA # 76226
datalivesite@gmail.com

*Abstract*— Personalised services bridge the gap between a financial institution and its customers and are built on trust. The more we trust the product, the keener we are to disclose our personal information in order to receive a highly personalized service that maximizes consumer value. Artificial Intelligence (AI) can help financial institutions tailor relevant products and services to their customers as well as improve their credit risk management, compliance, and fraud detection capabilities by incorporating chatbots and face recognition systems. Several works have been done in this area, however, none of the articles provides a comprehensive understanding of the research gaps and challenges in AI-based digital finance, which include explainability, trustworthiness, privacy, ethical consideration, and credit risk detection and mitigation. Addressing these research gaps and challenges will help to develop more transparent, trustworthy, ethical, and user-centric AI-based trust and customization tactics in digital banking. In the developing field of digital banking, these investigations will ultimately benefit financial service providers and consumers. The present study has analyzed sixteen research papers using the PRISMA model to perform a Systematic Literature Review (SLR). It has identified five research gaps and corresponding questions to analyze the present scenario. One of the gaps is credit risk detection for improved personalization and trust. Finally, an AI-based credit risk detection model has been built using four supervised machine learning classifiers viz., Support Vector Machine, Random Forest, Decision Tree, and Logistic Regression. Performance comparison shows an optimal performance of the model giving accuracy of ~89%, precision of ~88%, recall of ~89%, specificity of ~89%, F1_score of ~88%, and AUC of 0.77 for the Random Forest classifier. This model is foreseen to be most suitable for envisaging customer characteristics for which personalized credit risk mitigation strategies are particularly effective as compared to other existing works presented in this study.

## I. INTRODUCTION

Personalization and trust are intertwined aspects of a successful business-customer relationship. Personalization enhances the customer experience and engagement, while trust underpins customer confidence and loyalty [1]. Together, they create a strong foundation for long-term customer relationships and business success. Businesses that prioritize both personalization and trust are more likely to thrive in competitive markets. In the financial industry, digital finance is a sector that leverages new technology often known as FinTech (Financial Technology) to improve and automate the delivery and use of financial services. With net banking and credit card swipes for financial transactions, digital finance had its start a long time ago. However, because of the COVID-19 pandemic and the need for people to rely solely on digital payment methods while they were isolated, the stress on the entire digital banking system has increased significantly. Digital payments have increased dramatically after the COVID-19 pandemic.

However, FinTech and digital finance are not synonymous. All financial services and products that are enhanced or provided by digital technologies fall under the umbrella term of "digital finance." This covers everything, from blockchain-based solutions to automated investing platforms and mobile payments to online banking [2]. It is not exclusive to recent arrivals in the finance sector or startups in the technology sector. By embracing and incorporating digital technology into their current operations, traditional financial institutions like banks and insurance firms can also participate in the world of digital finance. It covers a broad range of financial services, including wealth management, lending, insurance, digital payments, and online banking. Its main goal is to leverage digital channels to increase the efficiency, convenience, and accessibility of these services. Like traditional financial services, digital finance is governed by financial regulations and oversight. Compliance with regulatory requirements is crucial, and digital finance providers must adhere to the same legal and security standards.

FinTech, as the name suggests, is a specific subsector of digital finance that focuses on technology-driven innovations in financial services. It typically involves technology startups and companies that disrupt and challenge traditional financial institutions. The specialized software and algorithms used by FinTech organizations are installed on PCs and smartphones. FinTech is essentially used to assist organizations, entrepreneurs, and consumers in managing their financial operations, workflows, and personal finances more effectively. These FinTech companies specialize in developing and offering innovative financial products and services that utilize digital technology, data analytics, artificial intelligence (AI), and automation to improve various aspects of financial transactions, management, customer satisfaction, trust, and decision-making. FinTech includes payment processing, online banking, facilitating loans and crowdfunding, enabling automated investment and wealth management, buying, selling, and trading digital currencies, development and use of cryptocurrencies, such as Bitcoin insurance purchasing, underwriting, and claims processing to name a few. Companies in the finance industry that use FinTech have expanded financial inclusion and use technology to cut down on operational costs. FinTech companies must also adhere to financial regulations, but they sometimes face unique regulatory challenges due to their innovative nature. Regulators often need to adapt to oversee new and unconventional FinTech services effectively.

The combination of personalization and trust management in digital finance creates a more engaging and secure environment for customers. When customers see that their unique needs and preferences are valued and that their data is handled with care and integrity, they are more likely to trust digital finance providers and develop long-lasting relationships with them. Trust is crucial in an industry where handling sensitive financial data and transactions is central to the business. In the banking industry, customer loyalty is a hard-earned trait. A bank's capacity to satisfy customers' needs is significantly more important than its reputation. People move on when their wants are not completely satisfied, frequently opening several bank accounts to meet their various needs. Statista conducted a poll with over 75,000 bank customers in 32 global locations to determine the most crucial elements to consider when selecting a bank. The report also provides country profiles that provide a detailed understanding of the degree of satisfaction with banking services. Figure 1 shows the bank customer trust score worldwide in 2023 by country. The figure is a bar plot that depicts the top 33 country's trust score with Indonesia leading the chart with a score of 4.38 followed by the Philippines (4.3) Brazil (4.28), India (4.23), and Saudi Arabia/South Africa (4.23) in the top five categories [3].

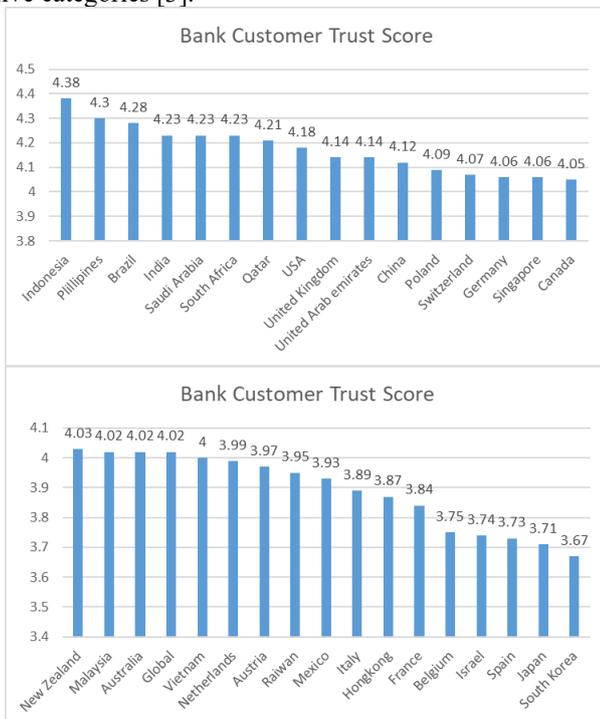

Fig.1. Bank Customer Trust Score Worldwide in 2023 by Country

Internet banking comes with a number of concerns. The evolving financial landscape poses a significant threat to data privacy, which banks must address, as third-party payment methods may result in the disclosure of consumer information. Bank management needs to be aware of the constantly evolving digital techniques that give rise to new hazards when it comes to automated operations. The established policies and procedures of the bank should take into account the growing demand from customers for adequate testing procedures and more stringent system checks to prevent any revenue leakage.

AI is a very promising and emerging area of research and development that may be employed in the digital finance sector for enhanced customer satisfaction, business relations, personalized portfolio optimization, risk assessment in real-time forex trading, health insurance products, personalized investment recommendations, personalized finance management for Small to Medium-sized Enterprises (SMEs), etc. There are several ways by which AI has revolutionized digital payments. Some of them, like Face-pay, use AI-powered face recognition technology, while others take a more futuristic approach, allowing users to be recognized by their smartphone app—like Moby in Shanghai—for smooth, error-free payments without the need for a PIN. Some payment gateways prevent fraudulent transactions based on artificial intelligence. Artificial intelligence (AI)-driven chatbots, also known as virtual assistants, can assist users in maintaining their financial well-being by assisting them in paying off debt (student loans, credit card debt, etc.), improving the way they manage their savings, and offering an unparalleled customer experience by handling accounts and facilitating payments with little to no waiting time. Digital payment systems and AI payment gateways are two essential elements for fintech success. Sellers may enhance customer satisfaction, boost platform stickiness, and generate longer loan-to-value ratios (LTV) with a larger customer base by implementing frictionless, safe payment gateways that are backed by AI technologies.

This study has performed a Systematic Literature Review (SLR) to identify, map, and analyze research gaps and research questions. Finally, the design of a comprehensive AI-based credit risk detection model has been presented for personalization, trust, overall customer satisfaction, and business-customer relationship building in the digital finance space using a secondary dataset for digital finance credit risk.

The main contributions of the paper include:
1. To identify, map, and analyze research questions and research gaps based on SLR.
2. To explore various AI-based models in the digital finance sector with regard to personalization and trust.
3. To propose an architectural concept of an AI-based credit risk detection model in digital finance.
4. To present exploratory data analysis of the dataset utilized for the study followed by a supervised machine learning classification model for a trustworthy digital finance ecosystem.

This is how the remainder of the paper is structured. Section II presents SLR followed by related works in Section III. Section IV is devoted to the proposed AI-based credit risk detection model in the digital finance domain including datasets used, data preprocessing, and exploratory data analysis followed by binary classification for credit risk or credit no-risk. Section V highlights results and discussion followed by concluding remarks and future scopes in section VI.

## II. SYSTEMATIC LITERATURE REVIEW

Preferred Reporting Items for Systematic Reviews and Meta-Analyses (PRISMA), an evidence-based minimal set of items for reporting in systematic reviews and meta-analyses, has been employed to conduct systematic literature reviews (SLR) in the present corpus of research [4]. PRISMA can be used as a basis for reporting systematic reviews of other types of research, such as interventions, even though it is primarily focused on the reporting of reviews of randomized trials..

A. Methodology

The steps of SLR are as follows.

1) Research topic

Defining the research topic is the first stage. Next, is planning a precise and thorough search strategy to find materials that are pertinent to the investigation based on inclusion and exclusion criteria as well as keywords.

2) Resource identification

Phase two entails assessing the discovered studies using the inclusion and exclusion criteria. Using this approach, the entire texts of the studies that meet the selection criteria are examined after their titles and abstracts have been evaluated to determine which ones meet the requirements. A database search was conducted to analyze 102 articles from academic publications and industry whitepapers published between 2011 and 2023. The focus of the analysis was on the articles' benefits and drawbacks, the techniques or analysis tools employed, the study environment, and the relationships examined from the findings. IEEE Xplore, Google Scholar, Web of Science, ResearchGate, Science Citation Index, Taylor & Francis, and SpringerLink are a few examples of databases. At first, only the keywords, abstracts, and titles from 2011 to 2023 were screened. The defined search terms, "AI in finance," "FinTech," "Trust in banking," "Personalization in finance," "Digital finance," "Bank customer relationship," "Big Data in digital finance," "Machine learning in credit risk," "Machine learning in credit rating," etc., are the only factors that determine the search strategy. A total of 102 papers were chosen for analysis in this step.

3) Data collection and analysis

The third phase is to use a standardized data extraction form to collect data from each study. Included are specifics on the participants, treatments or exposures, results, sample size, and other pertinent data. By using the following inclusion and selection criteria, the research was narrowed down to (a) studies published between 2011 and 2023; (b) peer-reviewed papers; and (c) publications in reputable English-speaking journals.

4) Data synthesis

Depending on the study's design and the research question, the fourth step involves synthesizing the results of the selected studies. A statistical or qualitative synthesis may be required at this point. The construction of the data analysis flow diagram, as illustrated in Figure 2, was based on the PRISMA paradigm.

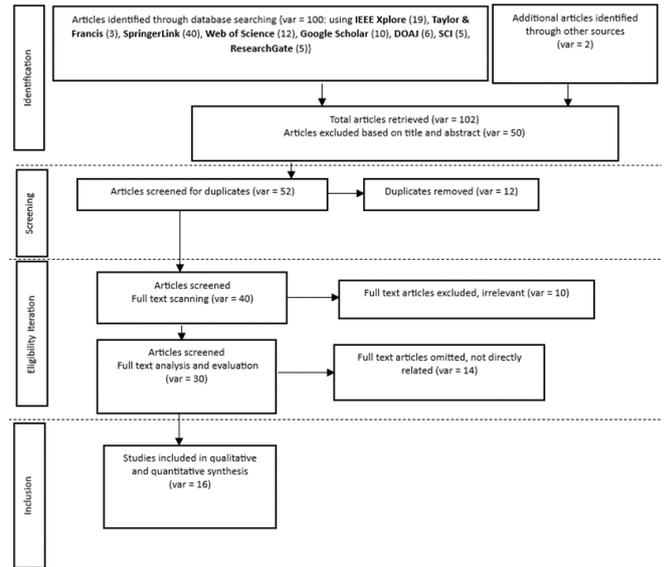

Fig. 2. PRISMA Flow Diagram

According to the graphic, 102 papers in total were found by utilizing the pertinent keywords. After a preliminary search to determine eligibility, 52 papers were found to be eligible. Online software was used to eliminate 12 duplicates. 40 papers were now selected for full-text scanning. Ten articles were removed because they had no bearing on the current investigation. Thirty papers remained for appraisal and full-text analysis. At least 14 irrelevant articles were deleted. In the end, 16 papers were selected for further analysis in order to conduct a qualitative and quantitative synthesis. These integers are represented by the variable 'var'.

B. Research gap analysis

Using the PRISMA methodology, the gaps in the literature were highlighted in this section. The research flow is shown in Table 1 with five research gaps (RG), corresponding research questions (RG), and a mapping that allows RQ-RG analysis to be performed. Identifying and filling these research gaps will assist in developing the understanding of trust and personalization in digital banking, leading to more effective, ethical, and consumer-centric financial services in the long run.

Table 1. Research flow

| RGs | RQs | RQ-RG analysis |
|---|---|---|
| Explainability and Trust in AI-Based Personalization | RQ1: How can AI-based personalization algorithms be made more transparent and explainable to enhance user trust in financial recommendations and decisions? | Transformed consumers' views of trust once they were able to decipher and comprehend the choices made by AI systems in the financial services industry |
| Trustworthiness and Data Privacy in AI-Driven Personalization | RQ2: What are the best practices for ensuring the security of user data in AI-driven personalization without compromising trust in the system? | Role of data protection regulations like GDPR and CCPA for shaping trust in AI-driven personalization in different regions |
| Ethical Considerations in AI-Based Personalization | RQ3: What ethical considerations are unique to AI-driven personalization in digital finance, and | Aligning AI-based personalization with ethical principles to enhance user trust in financial institutions |

| | how can they be addressed to build trust with users? | |
|---|---|---|
| Bias and Fairness in AI-Driven Personalization | RQ4: What are the potential biases that may arise in AI-driven personalization algorithms in digital finance, and how do they impact trust among different demographic groups? | Role of diversity and inclusion in building trust through AI-driven personalization, and their measurement and improvement |
| Personalization Strategies and Credit Risk Detection | RQ5: How can personalized financial advice and products be tailored to individual borrowers to enhance their ability to manage credit risk? | Customer segments or characteristics for which personalized credit risk mitigation strategies are particularly effective |

### III. RELATED WORKS

In Section II, the PRISMA model was used to select sixteen research papers to make a quantitative and qualitative analysis of AI models in finance. Using publicly available datasets, a novel classification methodology for machine learning (ML)-driven credit risk algorithms was proposed in [5]. They also demonstrated that, when it comes to credit risk estimation, the majority of deep learning models outperform traditional machine learning and statistical algorithms, and that ensemble methods yield a higher level of accuracy when estimating the performance ranking of individual models. Real-time risk assessment is a big difficulty in the financial business. In [6], different risk models with deep learning methods have been discussed. Systems for recommending investments and financial assets that use several recommendation algorithm types are in development [7]. Personalized portfolio optimization using genetic algorithms can include "robo-advisory," which consists of risk assessment and digital onboarding procedures that allow customers to be categorized and mapped to suitable risk levels [8].

Small and medium-sized enterprises (SMEs) and financial institutions may find greater efficiency in a system that provides automation and customization. Through a better understanding of their operations and financial situation, these are probably going to empower SMEs and promote the use of data in decision-making. Financial institutions can use all of the data at their disposal to provide SMEs with individualized value-added services on top of their main business [9]. Using screening techniques, transaction data from various sources is processed, analyzed, and enhanced with extra external data relevant to counterterrorism and anti-money laundering. Using machine learning algorithms, these technologies identify anomalous transaction patterns and linkages among aggregated data that point to typologies and dangers of money laundering or terrorist funding at the level of particular financial institutions [10].

Products powered by AI that provide individualized health insurance services are now in development. Utilizing the generated models, risk assessment and individualized coaching services are offered [11]. It is fashionable to use state-of-the-art ML/DL technologies and build a precise AI-powered model to infer various driving profiles. Insurance firms can utilize this driving profile model to better predict the risks associated with driving an insured vehicle and customize the services they offer [12]. The first version of "The IEEE Finance Playbook," which offers a roadmap for Trusted Data and Artificial Intelligence Systems (AIS) for Financial Services, was created by more than fifty industry thought leaders from banks, credit unions, pension funds, law firms, academia, and technology services organizations based out of Canada, the US, and the UK in response to the need to bring AI ethical practices to applications in the financial services industry [13]. Financial services technologists have been encouraged to prioritize ethical considerations and human well-being when applying data related to Artificial Intelligence Systems (AIS) through the use of an industry-specific playbook [14].

The writers of [15] go over a variety of banking methods and use cases, including the AI projects that Indian banks are pursuing. Financial crime and compliance management, customer insight and relationship management, credit risk, and customer service are key areas of concentration for the use of AI/ML in banking. [16] presents the specific applications and impacts of financial intelligence applications in the financial area from four aspects: fundamental operation, intelligent processing, data statistics, and risk monitoring. It does this by using an actual financial robot employed by a corporation as an example. Supporting a proactive regulatory approach before any financial harm occurs is the best way to ensure a sustainable future for AI innovation in the financial sector. This proactive strategy should put into effect sensible policies that represent jurisdiction-specific guidelines in accordance with carefully interpreted international norms [17].

[18] examines the practical importance of elucidating the basic theoretical questions of SupTech, which, from the viewpoint of supervisory authorities, is the use of technology to facilitate and improve supervisory operations. AI can be used in a variety of ways, such as personalization, to enhance user interface (UI) design. AI may tailor the UI design for each user by examining their behavior, preferences, and previous interactions. Providing pertinent content and cutting down on the amount of time it takes to find what users need, can enhance user experience [19]. The recognition and validation of user trust in AI-enabled systems has grown in importance in order to promote their adoption. A fundamental tenet of the human-computer interaction (HCL) field has been proposed: AI-enabled systems should adopt a more human-centric approach rather than a technological one [20].

None of these articles provides a complete understanding of the research gaps and challenges in AI-based digital finance, including explainability, trustworthiness, privacy, ethical consideration, and credit risk detection and mitigation based on AI. The present study has considered several gaps and has modeled one of the RGs – credit risk detection in digital finance.

### IV. AI-BASED ARCHITECTURAL CONCEPT OF CREDIT RISK DETECTION IN DIGITAL FINANCE

#### A. Concept

In this section, we have considered one of the RGs as depicted in Table 1, and propose an AI-driven trust and personalization model in digital finance based on bank credit risk related to loans. The interplay between bank credit risk, personalization, and trust in digital banking is complex. Personalization has an impact on both credit risk assessment and risk mitigation, and it is built on the proper treatment of client data. In turn, trust can impact customer financial behavior and lower credit risk for banks. In digital finance, trust must be built through personalization and responsible data practices The important points to consider in this context are as follows.

*1) Credit risk and trust*

Credit risk is the danger that a borrower will not fulfill their financial obligations, such as returning a loan. Trust plays a crucial role in credit risk assessment and management. Financial institutions and banks need to have faith in loan repayment from borrowers, and borrowers need to have faith in the fairness and transparency of the terms offered by banks.

*2) Data and personalization*

In digital finance, personalization is based on data collection and analysis to tailor offers and financial services to individual customers. Banks, for instance, might use customization to provide customers with unique credit limits, interest rates, and lending possibilities.

*3) Risk assessment and personalization*

Risk assessment and mitigation can benefit from personalization. Banks can provide individualized risk management plans depending on the financial circumstances of each client thanks to personalization.

*4) Building trust through personalization*

Personalization can foster confidence between customers and banks. When a bank provides customized financial offers and services based on the demands and financial situation of its customers, those customers are more likely to believe that the bank is committed to their financial well-being. To feel at ease and confident when using digital Data security and privacy are intrinsically linked to personalization.

*5) Data privacy and trust*

Personalization and trust are inextricably tied to data privacy and security. Customers must have confidence that their financial information will be treated with care and used purely for their benefit.

*6) Trust and credit risk mitigation*

Reducing danger is another way to use trust. Because they feel the bank is trustworthy and equitable, customers who have faith in their banks are more likely to maintain good credit behavior.

Considering all the six points, a credit risk detection system for banks has been designed in this study, whose architecture is presented in section B.

*B. Architectural framework*

The AI-based architectural framework for credit risk detection is shown in Figure 3. Supervised learning models in machine learning (ML) are linked to learning algorithms that examine data for classification. Figure 3 shows the diagrammatic representation of the credit risk detection framework. It comprises two sections such as data preprocessing and classification. The raw "Credit Risk" dataset is fed to the data preprocessor where it is converted to numeric followed by normalization to scale the values between 0 and 1. Next, Exploratory Data Analysis (EDA) was performed on the normalized data for feature extraction. The output of the preprocessor is then fed to the ML classifier for detecting credit risk (1) or credit no-risk (0). The ML classifiers used for classification comprise four algorithms: Support Vector Machine (SVM), Random Forest, Decision Tree, and Logistic Regression. The dataset is divided into two parts: training and testing. After training, a model file is created that is fed to the recognition block to recognize the test data. The output is the detection of credit risk or credit no-risk.

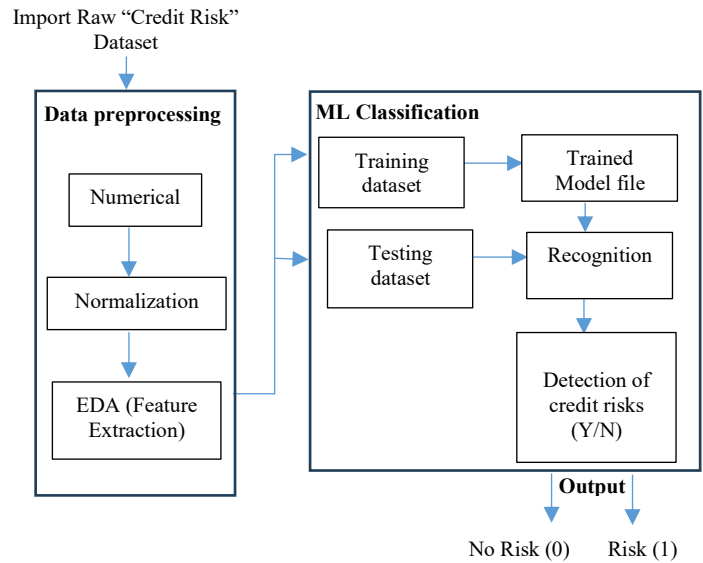

Fig. 3. ML-based credit risk detection framework

*C. Dataset*

The dataset that has been used to train the AI model is obtained from Kaggle [21]. The features used in the dataset comprise key factors affecting credit risk analysis. There are 12 fields comprising eleven features and one target {credit risk (1); credit no_risk (0)}. The features (numeric and categorical) are as follows:

1. Age: age of the borrower (numeric)
2. annual income: annual income of the borrower (numeric)
3. home ownership: own house, rented house, mortgage, others (categorical)
4. employment length (years): number of years of service of the borrower (numeric)
5. loan grade: grades A-G (categorical)
6. loan intent: debt consolidation, education, home improvement, medical, personal, venture (categorical)
7. loan amount: amount of loan borrowed (numeric)
8. interest rate: the loan interest rate (numeric)
9. percentage income: loan percent income (numeric)
10. history of default: whether the borrower has a previous loan non-repayment history
11. credit history length: 2 to 30 years (categorical)

The target variable is "loan status" (categorical).

*D. Exploratory data analysis*

The raw dataset has been analyzed by using various performance parameters simulated in Python such as heat map, correlation matrix, bar plot, box plot, pair plot, and distribution plot to access the characteristics of these features. Initial feature extraction out of the eleven features has been performed using the correlation matrix and the heat map of the features against the target variable as shown in Figure 4. The target variable gives (loan status – no default (0) i.e., no risk and default (1) i.e., credit risk). Feature extraction is depicted in Table 2. Features whose correlation (quality) values >±0.1 are included whereas those having correlation (quality) values <±0.1 are dropped. It is seen that four features were dropped. Seven features (F2, F3, F6, F7, F8, F9, F10) were included at this stage.

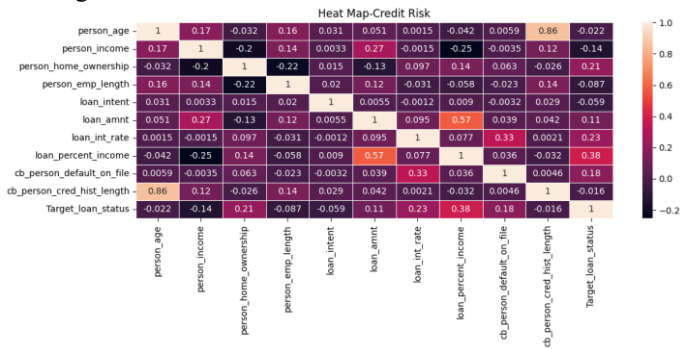

Fig. 4. Heat map of features with the target

Table 2. Feature extraction

| Feature Description | Feature Name to Include | Quality | Status |
|---|---|---|---|
| F1: Age | person_age | -0.021625 | Drop |
| F2: Annual income | person_income | -0.144551 | Include |
| F3: Homeownership | person_home_ownership | 0.211577 | Include |
| F4: Employment length (years) | person_emp_length | -0.087049 | Drop |
| F5: Loan intent | loan_intent | -0.058935 | Drop |
| F6: Loan grade | loan_grade | 0.373145 | Include |
| F7: Loan amount | loan_amnt | 0.105407 | Include |
| F8: Interest rate | loan_int_rate | 0.226825 | Include |
| F9: Percent income | loan_percent_income | 0.379371 | Include |
| F10: Historical default | cb_person_default_on_file | 0.179141 | Include |
| F11: Credit history length | cb_person_cred_hist_length | -0.015529 | Drop |
| F12: loan status | Target: no-risk (0), risk (1) | | |

Next, the characteristics of these features were examined by viewing the bar plots, box plots, distribution plots, and pair plots, which are presented in Table 3. Characteristics of features such as magnitude credit no-risk, magnitude risk, data type, and reliability are obtained from the bar plots. The magnitudes are depicted in the bar plots directly, whereas the data type can be observed to be balanced or unbalanced by checking the relative height of the bars. If the heights of the bars for both the target values of 0 and 1 are equal or nearly equal (equality>60%), a balanced data type is considered. Otherwise, unbalanced. Reliability can be found in the black lines towards the middle of the bars. These are called error bars. A short error bar or absence of error bar indicates reliable data whereas long error bars indicate unreliability in the data. Outliers, spread credit no-risk, and spread credit risk are directly obtained from the box plots. Box plots, which exhibit the data quartiles (or percentiles) and averages, visually represent the distribution of numerical data and skewness. Data points that do not fit the pattern are called outliers. The distribution plot resembles a histogram, only instead of bars, it displays a smooth curve. This plot's peak displays the highest concentration of numerical data. It examines how the values in the provided dataset are distributed visually. They are very helpful in aiding with the visualization of skewness. Variance is a metric for measuring data variation; the greater the variance, the more dispersed the data values are. If the distribution is regularly distributed, the skewness value is 0; if the skewness value is greater than 0, the distribution is left-skewed; if the skewness value is less than 0, the distribution is right-skewed. Values for skewness and variance shown in Table 3 are simulated in Python.

Table 3. Extracted feature characteristics

| Characteristics of extracted features | Extracted features Type of correlation (+ve/-ve) | | | | | | |
|---|---|---|---|---|---|---|---|
| | F2 (-ve) | F3 (+ve) | F6 (+ve) | F7 (+ve) | F8 (+ve) | F9 (+ve) | F10 (+ve) |
| Magnitude (credit no-risk/risk) | 0.019/0.007 | 0.5/0.78 | 0.16/0.34 | 0.25/0.34 | 0.23/0.36 | 0.17/0.29 | 0.0045/0.01 |
| Data type | UB | UB | UB | B | B | UB | UB |
| Reliability | high | high | high | high | high | high | high |
| Outliers | yes | yes | yes | yes | yes | yes | yes |
| Spread (credit no-risk/risk) | 0/0 | 1/0.35 | 0.3/0.3 | 0.2/0.25 | 0.3/0.3 | 0.15/0.25 | UD/0.03 |
| Variance | 0 | 0.228 | 0.0377 | 0.033 | 0.063 | 0.016 | 0 |
| Skewness | 32.8 (L) | -0.26 (R) | 0.86 (L) | 1.19 (L) | -0.63 (R) | 1.062 (L) | 1.69 (R) |
| Density peak (credit no-risk/risk) | 8 at 0/125 at 0 | 3 at 0 and 1/4.5 at 1 | 7 at 0/2.5 at 0.5 | 3,2 at 0.15/2.2 at 0.1 | 2.1 ay 0.1/2 at 0.6 | 4 at 0.1/2.5 at 0.4 | 250 at 0/100 at 0 |

Legend:
B: Balanced; UB: Unbalanced; N: Normal; UD: Undefined; R: Right Skewed (Positive Skewed); L: Left Skewed (negative Skewed).

Closely observing the feature values, it was noticed that the feature "F2: person_income" suffers from a few shortcomings such as oppositely unbalanced data type as compared to other features, very low magnitude for credit risk, zero spread, zero variance, unbalanced density peaks at target 0 and target 1. This is also clear from the pair plot. So, this feature was dropped. Finally, six features (F3, F6, F7, F8, F9, F10) were included for better classification fitment in the next part of the work.

V. SIMULATION AND PERFORMANCE ANALYSIS

A. Simulation environment

For simulation, the environment was set up for the classification of the final extracted dataset comprising six features (F3, F6, F7, F8, F9, F10) in Python. Classification was performed using four supervised ML classifiers for example Support Vector Machine (SVM), Random Forest, Decision Tree, and Logistic Regression. One kind of supervised learning algorithm that can be applied to tasks involving classification is the SVM. In the training data, SVM locates a hyperplane that maximally divides the various classes. Finding the hyperplane with the biggest margin—defined as the separation between the hyperplane and the nearest data points from each class—is how this is accomplished. New data can be categorized by identifying which side of the hyperplane it falls on once the hyperplane has been identified. SVMs are especially helpful in situations where the data has a large number of features and/or a distinct margin of separation. However, because they rely on intricate and nuanced algebraic concepts related to optimization theory, their implementation might be challenging [22].

Built on decision tree techniques, Random Forest is a supervised machine learning algorithm. This algorithm is used to forecast behavior and results in a variety of industries, including e-commerce and banking. The dichotomy rule fusion approach can be used to select a better feature sequence for Random Forest classification, increasing the classification accuracy. [23].

Among the most effective supervised learning techniques for classification applications is the decision tree. It constructs a tree structure that resembles a flowchart, with each internal node signifying a test on an attribute, each branch designating a test result, and each leaf node (terminal node) containing a class name. When a stopping criterion—such as the maximum depth of the tree or the minimum number of samples needed to split a node—is satisfied, the training data is recursively split into subsets based on the values of the attributes. It can be used to find data, extract text, locate missing data in a class, enhance search engines, and find additional uses including the medical field in place of statistical processes [24].

A supervised machine learning approach called logistic regression is mostly used for classification problems in which the objective is to predict the likelihood that an instance will belong to a particular class or not. This type of statistical method examines the connection between a group of independent factors and a set of binary dependent variables. It is an effective tool for determining if an email is spam or not, as well as for spotting credit card fraud [25].

For all the classifiers, 75% of the data have been used for training and 25% have been used for testing purposes. A comparative analysis of all the classifiers was made to get an estimate of the performances of using these algorithms for the model under study.

*B. Simulation Results*

Table 4 presents the Random Forest classifier's optimal performance in relation to other classifiers based on performance parameters, including accuracy, precision, recall, specificity, and F1-score. The number of true values and expected values for the four classifiers are shown in the confusion matrices as shown in Table 5. In the confusion matrix, shown in Figure 5, the number depicting credit no-risks identified as no-risk are called true negatives (TN); those depicting credit risks identified as credit risks are true positives (TP); those depicting credit risks identified as no-risks are called false negatives (FN); and the number depicting credit no-risks identified as risks are called false positives (FP). The proportion of accurately predicted instances to all instances is known as accuracy. The ratio of accurately anticipated positive instances to all expected positive instances is known as precision. The ratio of accurately predicted positive cases to the total number of positive instances that actually occurred is known as recall. This proportion of true positives in a binary classification is also known as the test's sensitivity recall. The percentage of true negatives that are successfully identified in a binary classification test is known as specificity. The harmonic mean of recall and precision is the F1-score.

| Actual No-risk | TN | FN |
|---|---|---|
| Actual Risk | FP | TP |
| | Predicted No-risk | Predicted Risk |

Fig. 5. Confusion matrix

The performance metrics are calculated based on the following equations.

$$\text{Accuracy} = \frac{TP+TN}{TP+TN+FP+FN} \quad (1)$$

$$\text{Precision} = \frac{TP}{TP+FP} \quad (2)$$

$$\text{Recall} = \frac{TP}{TP+FN} \quad (3)$$

$$\text{Specificity} = \frac{TN}{TN+TP} \quad (4)$$

$$\text{F1-score} = \frac{2*Precision*Recall}{Precision+Recall} \quad (5)$$

Plotting the rate of true positives versus false positives yields the Receiver Operating Characteristic (ROC) curve, which is used to assess the effectiveness of a classification model. The probability that a classifier would score a randomly selected positive instance higher than a randomly selected negative instance is known as the area under the ROC curve, or AUC. The value of AUC ranges from 0 to 1. When a model's AUC is 1, it can accurately categorize observations into classes; when it is zero, it is no more effective than a random guessing model. In machine learning, the $R^2$ score is a measure of how well a model fits data. It is a statistical measure of how close the data are to the model's predictions. The $R^2$ score is also called the coefficient of determination. The $R^2$ score ranges from 0 to 1, and the higher the score, the better the model fits the data. A score of 0 means that the model does not explain any of the variance in the data, and a score of 1 means that the model explains all of the variance. The $R^2$ score and ROC-AUC curves for the classifiers are presented in Figure 6.

Table 4. Performance metrics

| Classifier | Accuracy | Precision | Recall | Specificity | F1-score |
|---|---|---|---|---|---|
| SVM | 0.8853 | 0.8799 | 0.8853 | 0.8853 | 0.8778 |
| Random Forest | 0.8925 | 0.8889 | 0.8925 | 0.8925 | 0.8899 |
| Decision Tree | 0.8787 | 0.8737 | 0.8787 | 0.8787 | 0.8708 |
| Logistic Regression | 0.8397 | 0.8277 | 0.8397 | 0.8397 | 0.8258 |

Table 5. Confusion Matrix

| Confusion Matrix | Classifier | | | |
|---|---|---|---|---|
| | SVM | Random Forest | Decision Tree | Logistic Regression |
| | [[6252 230] [ 704 960]] | [[6144 338] [ 537 1127]] | [[6132 237] [ 751 1026]] | [[6040 329] [ 976 801]] |
| TP | 960 | 1127 | 1026 | 801 |
| TN | 6252 | 6144 | 6132 | 6040 |
| FP | 704 | 537 | 751 | 976 |
| FN | 230 | 338 | 237 | 329 |

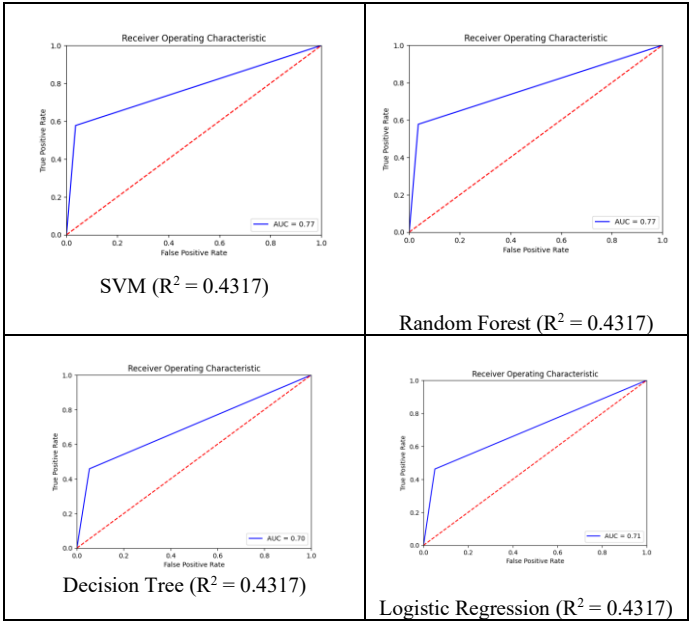
Fig. 6. ROC-AUC Curves

*C. Performance analysis and discussion*

A careful examination of the performance metrics shows the optimal performance of the Random Forest classifier with a performance index of ~89%, AUC of 0.77, and an R2 score of 0.4317. This is closely followed by SVM showing ~88% performance index, AUC of 0.77, and an R2 score of 0.4317. The decision Tree has a performance index of ~87%, an AUC of 0.70, and an R2 score of 0.4162 whereas, Logistic Regression has a performance index of ~83%, AUC of 0.71, and an R2 score of 0.4317. Sometimes moderate R2 values are obtained in any field that attempts to predict human behavior, like this model. The moderate R2 value but statistically significant predictors and high-performance metrics in the present credit risk detection model designed in this study are expected to give the best result with the Random Forest classifier.

A comparative analysis of the present model with other ML models related to credit risk datasets proves the enhanced performance of the present model with respect to accuracy and AUC as depicted in Table 6 [5].

Table 6. Accuracy of credit risk datasets

| Dataset | Model | Accuracy |
|---|---|---|
| Balanced FICO (Fair Issac Corporation) | Two-Layer Additive Risk Model | 0.7404 |
| 76 small businesses from a bank in Italy | Feedforward networks | 0.87 |
| NYU's Salomon Center database | Boosting | 0.8631 |
| A leading European p2p Platform Bondora | Neural networks | 0.7438 |
| An original credit scoring dataset of a Singaporean firm providing credit and loans | Neural networks | 0.84 |
| A real-world p2p Chinese data platform | AM LSTM | 0.669 (Auc) |
| Lending Club | A denoising auto-encoder-based neural network model | 0.875 |
| A real-world Chinese dataset | Relief-CNN | 0.6989 (Auc) |
| The present model under study using a Credit risk dataset from Kaggle | 4 Supervised machine learning models | 0.8925 0.77 (Auc) |

## VI. CONCLUSION

The present study has made the SLR of AI-based trust and personalization in the digital finance ecosystem using the PRISMA model. Five RGs have been identified and corresponding RQs have been formulated to make an RQ-RG analysis. The study further investigates one of the RQs and has used an ML-based credit risk detection model to detect credit risks thereby enhancing trust among the customers and the financial institutions. Using EDA, the present study has extracted six feature vectors that have been fed to four ML classifiers to detect risk. The simulation result analysis using the Random Forest classifier shows a performance of ~89% accuracy, ~88% precision, ~89% recall, ~89% specificity, ~88% F1 score, and 0.77 AUC, which is the best among the four ML classifiers used in this study. Further comparison with existing related works proves its effectiveness with respect to accuracy and AUC. This makes the model most suitable for detecting customer characteristics for which personalized credit risk mitigation strategies may be adopted in the future that are particularly effective. This model will enable personalized financial advice and products to be tailored to individual borrowers to enhance their ability to manage credit risk.

This model has provided a solution to one of the five RGs. There is a future plan to provide complete solutions to the remaining four RQs by designing related improved AI algorithms and Explainable AI (XAI) models.